\documentclass[a4paper,11pt]{article}
\usepackage{jheppub} % for details on the use of the package, please see the JINST-author-manual
\usepackage{lineno}
\usepackage{orcidlink}
\usepackage{etoolbox}

\title{\boldmath Topological defect induced phase separation in a holographic system}

\author[a]{Zi-Qiang Zhao\orcidlink{0009-0009-7859-3655},}%\email{zhaoziqiang@stumail.neu.edu.cn}
\author[b,\ast]{Zhang-Yu Nie\orcidlink{0000-0001-7064-247X},}
\author[a]{Jing-Fei Zhang\orcidlink{0000-0002-3512-2804}}%\email{jfzhang@mail.neu.edu.cn}
\author[a,c,d,\ast]{and Xin Zhang\orcidlink{0000-0002-6029-1933}\note[$\ast$]{Corresponding author.}}

\affiliation[a]{Liaoning Key Laboratory of Cosmology and Astrophysics, College of Sciences, Northeastern University, Shenyang 110819, China}
\affiliation[b]{Center for Gravitation and Astrophysics, Kunming University of Science and Technology, Kunming 650500, China}
\affiliation[c]{MOE Key Laboratory of Data Analytics and Optimization for Smart Industry, Northeastern University, Shenyang 110819, China}
\affiliation[d]{National Frontiers Science Center for Industrial Intelligence and Systems Optimization, Northeastern University, Shenyang 110819, China}

\emailAdd{zhaoziqiang@stumail.neu.edu.cn}
\emailAdd{niezy@kust.edu.cn}
\emailAdd{jfzhang@neu.edu.cn}
\emailAdd{zhangxin@neu.edu.cn}

\abstract{We investigate the coupled dynamics of symmetry breaking and phase separation during quenches across the critical point in a first-order phase transition. Based on the Einstein-Maxwell-scalar theory, we construct a holographic superfluid model with $\mathbb{Z}_2$ symmetry. By introducing higher-order nonlinear terms $\lambda\Psi^4$ and $\tau\Psi^6$ into the scalar field potential, we realize a rich phase structure, which enables us to study the coupling effects between symmetry breaking and phase separation. Furthermore, by preparing initial conditions with well-defined spatial partitions, we discover a new triggering mechanism for the invasion phenomenon, namely that kinks serve as triggering sites for the phase separation process. This study reveals a novel coupling mechanism between topological defects and phase separation, enriches our understanding of nonequilibrium structure formation in strongly coupled systems.}
\begin{document}
	\maketitle
	\flushbottom
	
	\section{Introduction}
	\label{sec:intro}
	%非平衡动力学过程是物理学研究的重要前沿之一，其核心在于理解系统在外部驱动下如何从非平衡态弛豫至平衡态，以及在此过程中空间结构如何自发形成。全息对偶（AdS/CFT对偶）为研究强耦合量子多体系统的此类过程提供了一个独特的理论框架，能够自然地处理强相互作用下的非线性演化。近年来，利用全息方法研究相变动力学、拓扑缺陷生成以及非平衡结构形成，已成为领域内的热点方向。
	Nonequilibrium dynamical processes constitute one of the important frontiers in physics research. The central objective is to understand how systems relax from nonequilibrium states to equilibrium, and how spatial structures emerge spontaneously during this process. 
	Holographic duality (the AdS/CFT correspondence \cite{Maldacena:1997re}) provides a unique theoretical framework for studying such processes in strongly coupled systems \cite{Hartnoll:2008vx,Hartnoll:2008kx,Herzog:2010vz,
		Gubser:2008wv,Cai:2013aca,Chen:2010mk,Kim:2013oba,Basu:2010fa,Musso:2013ija,Nie:2013sda,Donos:2013woa,Li:2017wbi,Nie:2015zia,Nie:2014qma,Amado:2013lia,Alberte:2017oqx,Zhang:2021vwp,Zeng:2022hut,Baggioli:2022aft,Xia:2023pom,Xia:2021jzh,Zhao:2023ffs,Su:2023vqa}. 
	In recent years, the use of holographic methods to investigate phase transition dynamics \cite{Chen:2022cwi,Li:2020ayr,Chen:2022tfy}, topological defect formation \cite{Xia:2020cjl,delCampo:2021rak,Li:2021iph,Li:2021dwp,Xia:2021xap,Zeng:2022hut,Yang:2025bsw,Xia:2026yrj}, and nonequilibrium structure formation \cite{Xia:2023pom,Su:2023vqa,Zhao:2023ffs,An:2024ebg,Yang:2024hom,Xia:2024ton,Zeng:2024rwn} has emerged as a prominent research direction in the field.

	%当系统被快速驱动穿过临界点时，其演化行为通常由对称性破缺机制主导。对于具有$\mathbb{Z}_2$对称性的系统，这一过程会导致体系从对称相自发选择两个简并基态之一，从而破坏原有的离散对称性。在不同选择区域的交界处，会形成拓扑缺陷——畴壁（kink）。这类缺陷的生成密度与驱动速率之间的关系由Kibble‑Zurek机制描述。另一方面，当系统处于一阶相变的不稳定区或亚稳区时，即使不跨越临界点，均匀态也会因热力学失稳而自发分拆为具有不同序参量值的空间区域。这一现象被称为相分离，其典型特征是通过旋节线分解（spinodal decomposition）或气泡成核生长形成非均匀空间结构。
	When a system is rapidly driven across a critical point, its evolution is typically governed by the mechanism of symmetry breaking. For a system with $\mathbb{Z}_2$ symmetry, this process causes the system to spontaneously choose one of two degenerate ground states from the symmetric point, thereby breaking the original discrete symmetry. At the interfaces between regions choosing different states, topological defects known as domain walls (kinks) are formed. The relationship between the density of such defects is described by the Kibble-Zurek (KZ) mechanism \cite{Hendry:1994ami,doi:10.1126/science.251.4999.1336,PhysRevLett.83.5030,PhysRevLett.81.3703,PhysRevLett.84.4966,PhysRevLett.112.035701,Ulm2013,Sadler2006,PhysRevLett.121.200601,PhysRevLett.124.240602,Xia:2020cjl,Xia:2021jzh,Zeng:2022hut,Xia:2023pom,Suzuki:2023hag,PhysRevLett.127.115701,Sonner:2014tca,PhysRevLett.128.135701,Xia:2026yrj}. On the other hand, when the system lies in the unstable or metastable region of a first-order phase transition, even without crossing the critical point, the homogeneous state can spontaneously separate into spatial regions with different order parameter values due to thermodynamic instability. This phenomenon is called phase separation \cite{Janik:2015iry,Janik:2016btb,Janik:2017ykj,Attems:2017ezz,Attems:2019yqn,Bellantuono:2019wbn,Attems:2020qkg,Chen:2022tfy,Ning:2023edr,Zhao:2023ffs}, whose hallmark is the formation of inhomogeneous spatial structures through spinodal decomposition or bubble nucleation and growth.

	%上述两种机制——对称性破缺与相分离——通常被分别研究。然而，在实际的复杂体系中，它们可能同时存在并相互耦合。当淬火路径同时满足两个条件——即既穿越临界点触发对称性破缺，又进入一阶相变的不稳定区域——系统将经历两种机制的协同作用。此时，一个自然的问题是：这两种机制如何相互影响？特别是，对称性破缺产生的拓扑缺陷是否会反过来影响相分离的演化路径？本文将主要集中于研究这个问题。
	The two mechanisms mentioned above, symmetry breaking and phase separation, are typically studied separately. However, in realistic complex systems, they may coexist and couple with each other. When the quench path simultaneously satisfies two conditions, namely crossing the critical point to trigger symmetry breaking and entering the unstable region of a first-order phase transition, the system experiences the synergistic effects of both mechanisms. The natural question then arises: how do these two mechanisms influence each other? In particular, do the topological defects generated by symmetry breaking in turn affect the evolution pathway of phase separation? This paper will focus on investigating this issue.

	%本文基于爱因斯坦-麦克斯韦-标量理论，研究了一个具有$\mathbb{Z}_2$对称性的全息超流体模型。通过在标量场势能中引入$\lambda\Psi^4$和$\tau\Psi^6$两类高阶非线性项，我们可以调控系统的相变类型，使其呈现从二级相变到一级相变乃至洞穴相变（COW）的丰富行为。该模型为在同一理论框架内同时研究对称性破缺、相分离及其耦合效应提供了理想的平台。
	This paper investigates a holographic superfluid model with \(\mathbb{Z}_2\) symmetry based on the Einstein-Maxwell-scalar theory. By introducing higher-order nonlinear terms, namely \(\lambda\Psi^4\) and \(\tau\Psi^6\), into the scalar field potential, we can tune the type of phase transition in the system, enabling it to exhibit rich behaviors ranging from second-order to first-order and even Cave-of-Wind (COW) phase transitions \cite{Zhang:2021vwp,Zhao:2022jvs,Zhao:2024jhs,Zhao:2025tqq,Zhao:2025vtr,Wang:2025ajn}. This model provides an ideal platform for simultaneously investigating symmetry breaking, phase separation, and their coupled effects within a single theoretical framework. %Furthermore, this model allows for the convenient realization of supercritical phenomena \cite{10.1063/PT.3.1796}. When the control parameter of a system crosses the critical point, the first-order phase transition disappears, and the system enters a region where the thermodynamic properties vary continuously, yet distinct thermodynamical and dynamical behaviors can still be identified. Recent studies have shown that the supercritical region is not completely homogeneous and indistinguishable, and different subregions can still be delineated through dynamical methods \cite{Yoon_2018,PhysRevLett.111.145901,Bolmatov2013,Prescher_2017,Bolmatov_2015,Fomin_2018,Fomin2015,PhysRevE.85.031203,2023PhRvR...5a3149H,jiang2024experimental} or the analysis of thermodynamic response functions\cite{Xu_2005,Ruppeiner_2012,PhysRevLett.112.135701,PhysRevE.95.052120,Gallo2014}. Moreover, supercritical phenomena have also become a highly active research topic in black hole thermodynamics \cite{{Zhao:2025ecg,Xu:2025jrk,Li:2025tdd,Wang:2025ctk,Li:2025lrq,Anand:2025rzh,Guo:2026xlk}}. 
	
	We focus on analyzing how the system undergoes a coupled evolution of symmetry breaking and phase separation during a quench that simultaneously crosses the critical point and enters the unstable region of a first-order phase transition. By introducing initial conditions with well-defined spatial partitions, we discover a novel dynamical phenomenon: topological defects can serve as preferential triggering sites for phase separation, inducing a directional ``invasion'' process whose propagation velocity exhibits spatial scale independence under ultrafast quenches. This finding reveals a new mechanism for the coupling between defects and phase separation. %Furthermore, we employ this velocity as a probe to investigate its behavior in the supercritical region, and find that the invasion velocity as a function of charge density exhibits a distinct turning point. This turning point defines a dynamical crossover lin in the phase diagram, providing a new criterion for distinguishing different subphases within the supercritical region.
	
	The remainder of this paper will be organized as follows. In section \ref{sec2}, we will introduce the holographic model and the parameter space of nonlinear terms. In section \ref{sec3}, we will describe the dynamical setup and the full nonequilibrium evolution. Finally, we will provide a summary in section \ref{sec5}.
	
	%这篇文章剩下的部分可以总结成如下章节。在第二章节，我们将介绍全息模型和非线性项的参数空间。在第三章节，我们将介绍动态程序和完整的非平衡演化。在第四章节，我们将介绍我们定义出来的新的超临界跨越的曲线。最后我们在第章节给出一些总结。

	\section{Holographic superfluid in Einstein-Maxwell scalar theory}\label{sec2}
	\subsection{Setup}
	In this paper, we consider the Einstein-Maxwell scalar theory, where the scalar field is neutral, implying that the system possesses a global $\mathbb{Z}_2$ symmetry. In addition, we include two higher-order nonlinear terms $\lambda\Psi^4$ and $\tau\Psi^6$, which are introduced to induce a first-order phase transition. The Lagrangian of our holographic system takes the following form
	\begin{align}
		\mathcal{L}_{m}=&-\frac{1}{4}h(\Psi)F_{\mu\nu}F^{\mu\nu}-\nabla_{\mu}\Psi \nabla^{\mu}\Psi-m^{2}\Psi^2-\lambda\Psi^4-\tau\Psi^6,\label{Lag}
	\end{align}
	in which $h(\Psi)=e^{\alpha \Psi^2}$. $\Psi$ is an uncharged scalar field and $F_{\mu\nu}=\nabla_{\mu}A_{\nu}-\nabla_{\nu}A_{\mu}$ is the Maxwell field strength. For the static solution, we adopt the ansatz of the following form
	\begin{align}\label{ansatz}
		\Psi=z\psi(z)/L~, A_\mu dx^\mu=\phi(z)dt~,
	\end{align}
	and the metric of the black hole is given by
	\begin{align}
		ds^{2}=\frac{L^2}{z^2}(-f(z)dt^{2}+\frac{1}{f(z)}dz^{2}+dx^{2}+dy^{2}),
	\end{align}
	where the function $f(z)$ is given by
	\begin{align}
		f(r)=1-(z/z_h)^3~,
	\end{align}
	where $z_h$ is the radius of the black hole event horizon, and its Hawking temperature is
	\begin{align}
		T= \frac{3}{4\pi z_h}.
	\end{align}
	The equations of motion are
	\begin{align}
		\psi  \left(\frac{f'}{f z}-\frac{m^2}{f z^2}+\frac{\alpha  z^2 \phi '^2
			e^{\alpha  \psi ^2 z^2}}{2 f}-\frac{2}{z^2}\right)+\frac{f' \psi '}{f}-\frac{2 \lambda  \psi
			^3}{f}-\frac{3 \tau  \psi ^5 z^2}{f}+\psi ''=0~,\\
		2 \alpha  \psi  z^2 \psi ' \phi '+2 \alpha  \psi ^2 z \phi '+\phi ''=0~.
	\end{align}
	To solve these differential equations, we need to specify the boundary conditions. The boundary condition at infinity is given as
	\begin{align}
		\phi(z)= \mu-z\rho+\dots~,~\psi(z)= \psi^{(1)}+ \psi^{(2)}+\dots~.
	\end{align}
	At the horizon, we impose $\psi(z_h)=0$. In the remainder of this paper, we work in the canonical ensemble, which means that the total charge $\rho$ is held fixed. Moreover, we choose the boundary condition with a vanishing source, which implies $\psi^{(1)}=0$. In this case, the nonzero vacuum expectation value is $\langle \mathcal{O}\rangle =\psi^{(2)}\neq 0$.
	
	In this model, we can obtain various types of phase transitions, including zeroth-order, first-order, second-order, and COW phase transitions. The most important tool for determining which type of phase transition the system undergoes is the free energy. Therefore, we also need to provide the formula for the free energy. In the probe limit, where only the matter field contributes to the on-shell action, the free energy formula, after omitting the spatial volume, is as follows
	\begin{align}
		G=\frac{V_2}{T}\left(\frac{\mu\rho}{2L^2}+\int_{0}^{z_h}\left(\frac{1}{2}e^{z^2\alpha\psi^2}z^2\alpha\psi^2\phi'^2- \lambda\psi^4-2z^2\tau\psi^6\right)dz\right),
	\end{align}
	where $V_2$ is the volume of the spatial boundary manifold. In addition, we also set $L = 1$, $z_h = 1$, $m^2 = -2$, and $\alpha=5$.
	%此外，我们还将固定L=1，zh=1以及m^2=-2.
	
	\subsection{Static solutions and phase diagram}
	\begin{figure}
		\centering
		\includegraphics[width=0.45\columnwidth]{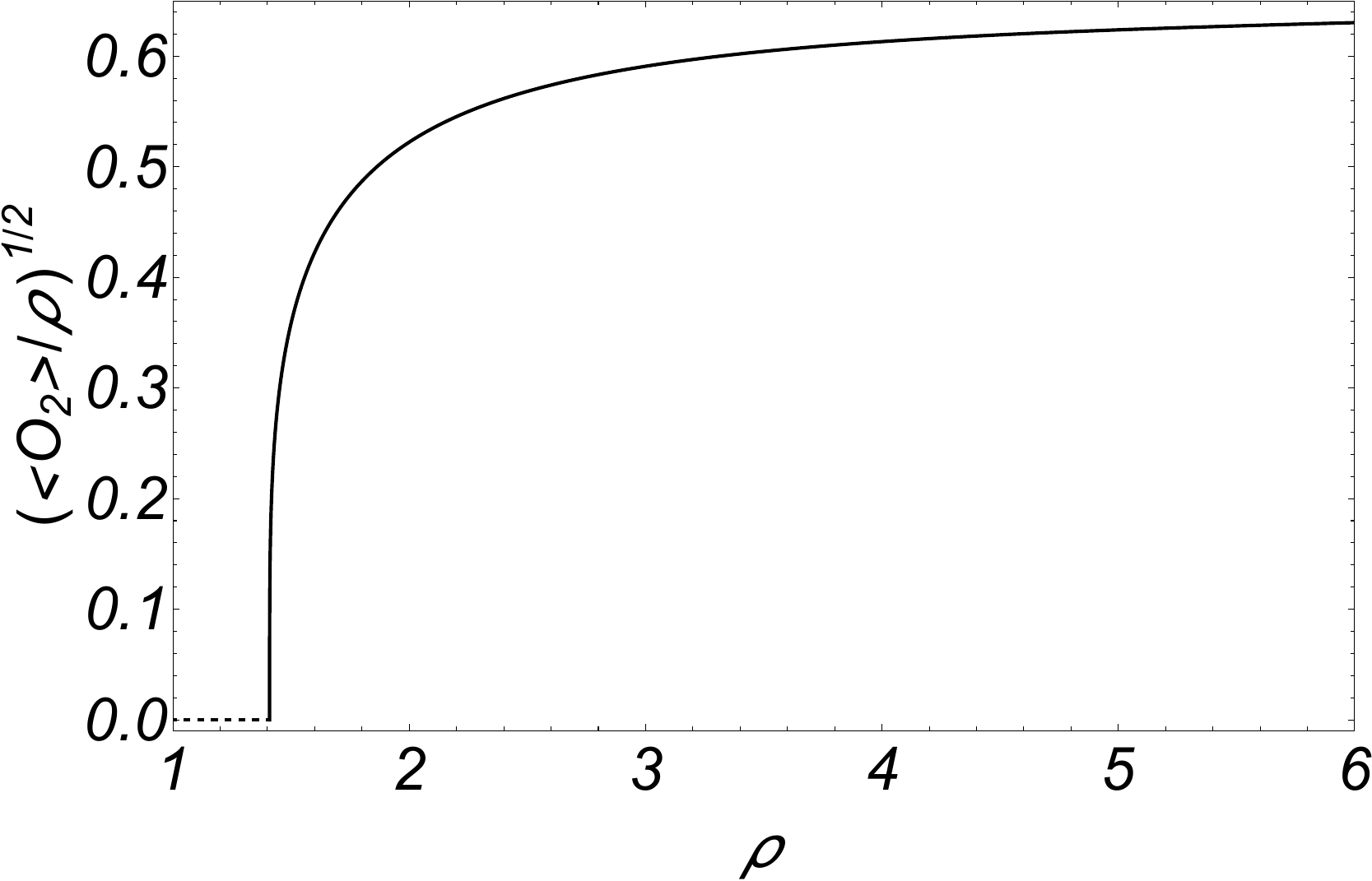}
		\includegraphics[width=0.45\columnwidth]{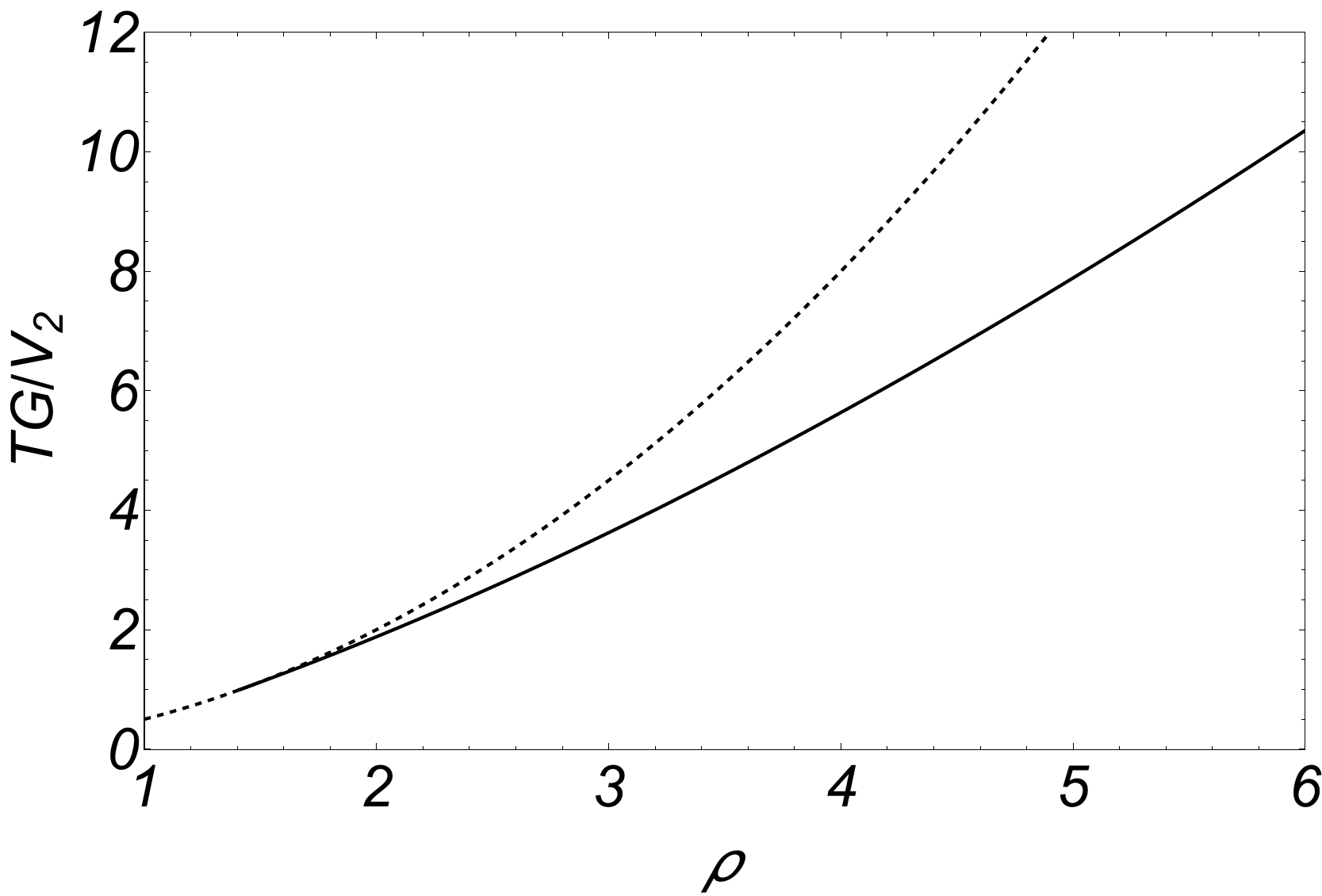}
		\caption{The condensate and free energy for $\lambda=0$ and $\tau=0$. The dashed lines correspond to the normal solution, and the solid lines correspond to the superfluid solution.}\label{co2nd}
	\end{figure}
	\begin{figure}
		\centering
		\includegraphics[width=0.45\columnwidth]{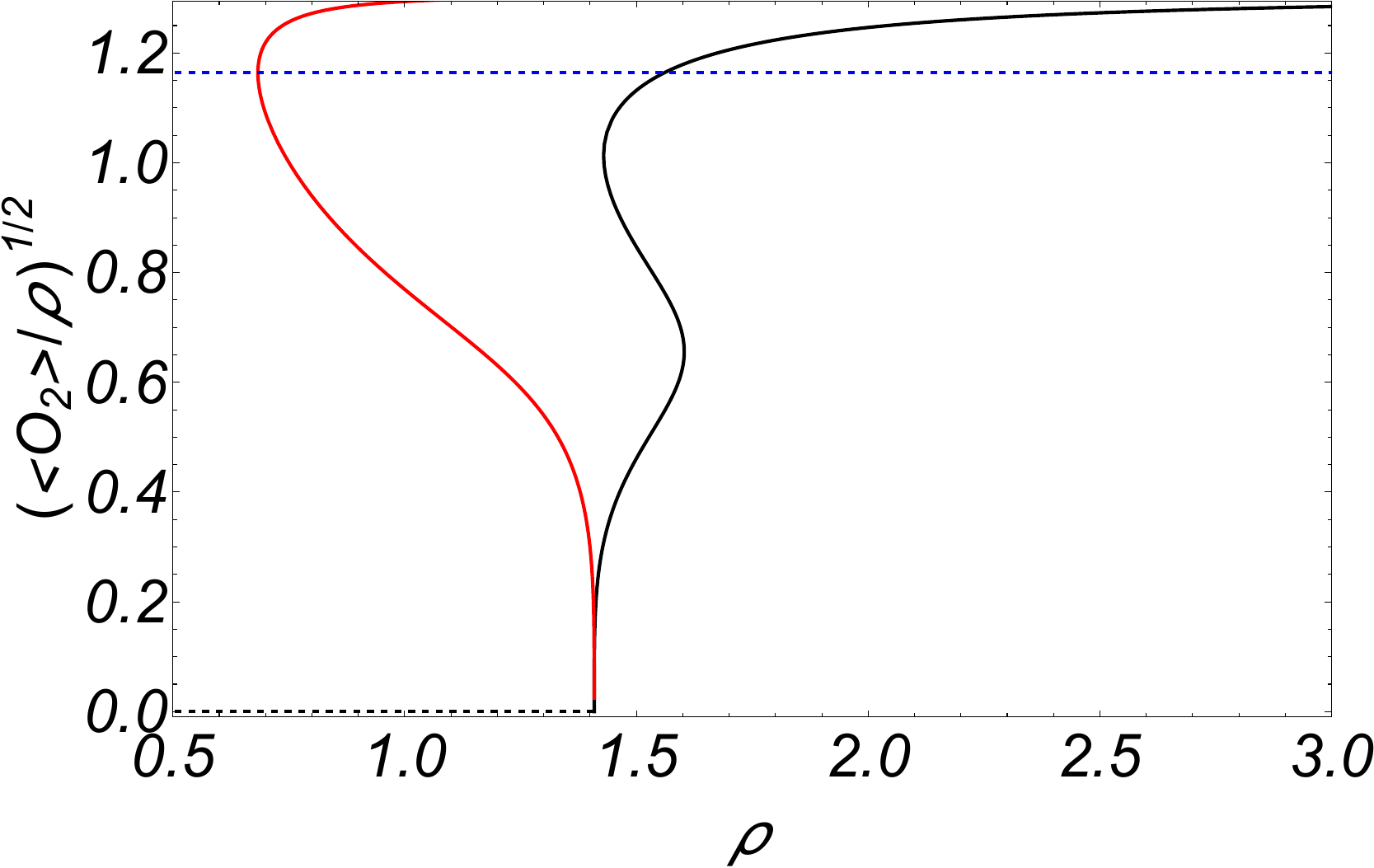}
		\includegraphics[width=0.45\columnwidth]{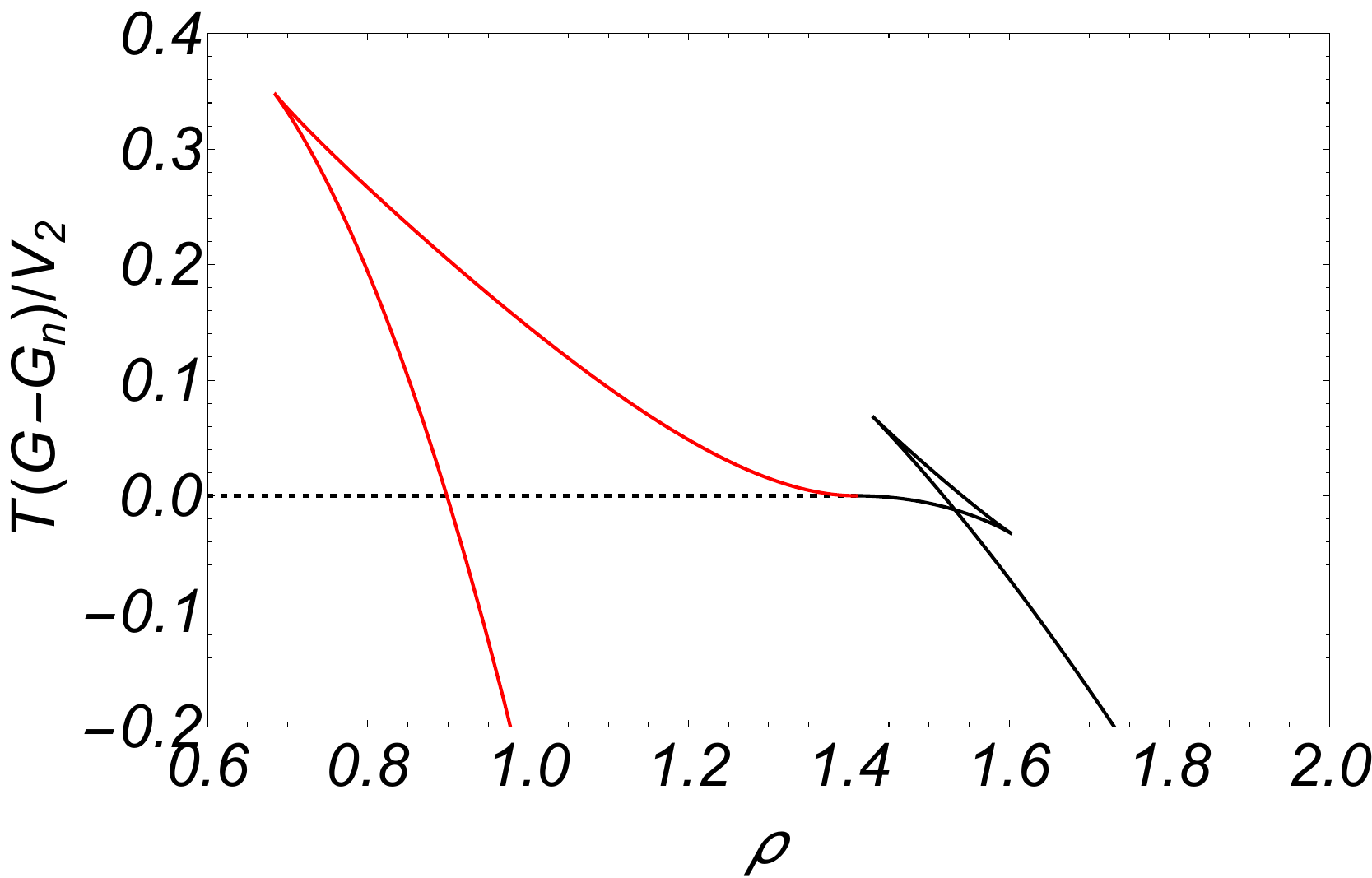}
		\caption{The condensate and free energy for $\lambda=-4$ and $\tau=2.7$. The black dashed lines correspond to the normal solution, and the black solid lines correspond to the superfluid solution. The red solid lines correspond to the grand canonical ensemble (fix chemical potential $\mu$) superfluid solution. The blue dashed line denotes the inflection point in the grand canonical ensemble. Where $G_n$ denotes the free energy of the normal solution.}\label{co1st}
	\end{figure}
	%蓝色虚线是mu系综的拐点
	
	In this model, when both $\lambda$ and $\tau$ are zero, the system reduces to the standard second-order phase transition, with the critical point $\rho_c \approx1.409$. We show the condensate and the corresponding free energy for the second-order phase transition in Fig.~\ref{co2nd}. In Ref.~\cite{Zhao:2022jvs}, the authors demonstrated that by introducing higher-order nonlinear terms, the system can exhibit various types of phase transitions. This is also the case in the neutral scalar field of the EMs model. 
	
\begin{figure}
		\centering
		\includegraphics[width=0.5\columnwidth]{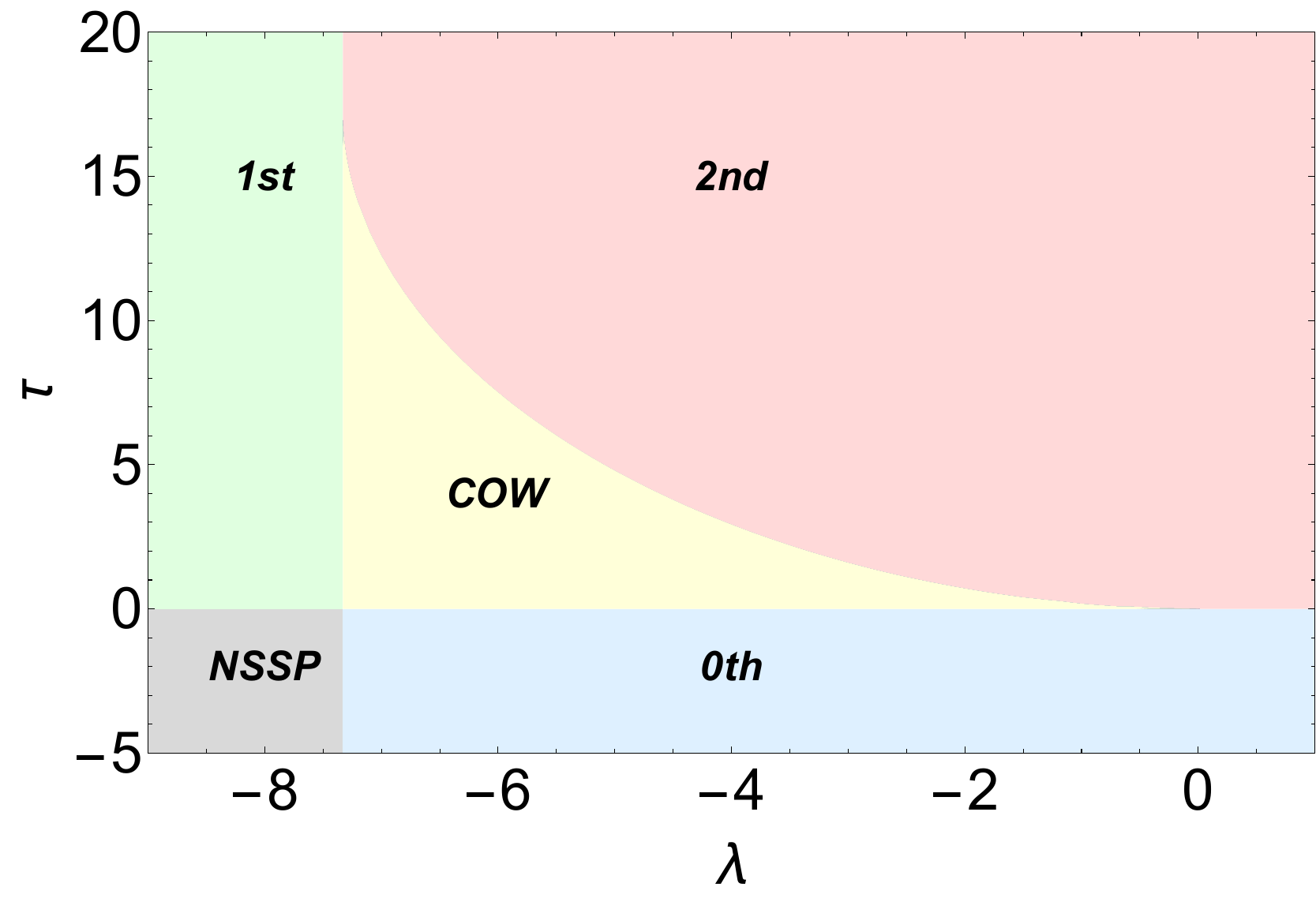}
		\caption{Phase diagram of the holographic system, where $NSSP$ denotes no stable superfluid phase.}\label{Phasediagram}
	\end{figure}

	%COW是一阶相变和二阶相变的组合.巨正则系综的自由能需要在正则系综自由能的基础上做一个勒让德变换。
	In Fig.~\ref{co1st}, we present the condensate and free energy for the COW phase transition. The COW phase transition represents a combination of first-order and second-order phase transitions, in which the system first undergoes a second-order phase transition to superfluid solution 1, and then a first-order phase transition from superfluid solution 1 to superfluid solution 2. In Fig.~\ref{co1st}, the black solid line represents the canonical ensemble, and the red solid line represents the grand canonical ensemble. In the grand canonical ensemble, the free energy is obtained from that of the canonical ensemble via a Legendre transformation. Once the spatial direction is turned on, the dynamical stability of the system is consistent with that of the grand canonical ensemble, so the phase separation region is clearly determined by the unstable region of the grand canonical ensemble (see Ref.~\cite{Zhao:2023ffs} for details). Under the parameters given in Fig.~\ref{co1st}, the region where the system develops inhomogeneous structures after being quenched from a stable solution to the spinodal region corresponds to the region where $\rho$ is less than the value at the intersection of the blue dashed line and the black solid line. This region corresponds to the unstable region in the grand canonical ensemble.
	In Fig.~\ref{Phasediagram}, we present the phase diagram for $\alpha=5$ in the $\lambda$-$\tau$ plane, which clearly delineates the parameter space corresponding to different phase transitions.
	%其中Gn表示的是normal解的自由能
	
	%此外，通过加高阶非线性项的方式，我们还可以很方便的在全息模型中实现超临界区域，文章[]中就通过这种方式研究过超临界区域。超临界区域不仅在微观和宏观中重要，在黑洞物理中也非常重要。
	It is worth noting that in the phase diagram presented here, we have only computed the curve from the COW phase transition to the second-order phase transition. According to the analysis in Ref.~\cite{Zhao:2022jvs}, in theory, as long as the parameter $\tau$ is nonzero, for sufficiently large $\psi$, one can always find a stable solution with lower free energy. This is also the case in our model. Furthermore, by introducing higher-order nonlinear terms, the supercritical region can be easily realized in the holographic model. This approach was used in Ref.~\cite{Zhao:2024jhs} to investigate the supercritical region. The supercritical region is significant not only in microscopic and macroscopic contexts but also in black hole physics \cite{Zhao:2025ecg,Xu:2025jrk,Li:2025tdd,Wang:2025ctk,Li:2025lrq,Anand:2025rzh,Guo:2026xlk}.

	\section{Symmetry breaking and phase separation}\label{sec3}
	In this section, we will introduce the complete dynamical evolution process. For quench dynamics across the critical point, this corresponds to a process of spontaneous symmetry breaking. However, during the process of symmetry breaking, due to differences in relaxation times, the system may generate topological defects. For a one-dimensional system with $\mathbb{Z}_2$ symmetry, the topological defects produced are kinks. The number of topological defects is described by the KZ mechanism. In addition, if the system exhibits instability (for example, in the presence of a first-order phase transition), inhomogeneous structures can be generated even during a quench process that does not cross the critical point. These inhomogeneous structures are called bubbles, and this process is referred to as phase separation. In the following section, we will introduce these two mechanisms, respectively, and their combined effects.
	
	% If we quench across the critical point into the unstable region of a first-order phase transition, the system will undergo not only symmetry breaking but also phase separation. A natural question arises: which of these two mechanisms will dominate the evolution of the system in this process? Furthermore, if these two mechanisms are combined, will some novel and peculiar physical phenomena emerge? In the following sections, we will address this question.
	%在接下来的这个章节中，我们将分别介绍这两种机制以及他们的混合效应。
	
	%非平衡动力学演化的方程
	\subsection{Equations for nonequilibrium dynamical evolution}
	For the dynamical process, we adopt the ansatz of the following form
	\begin{align}
		\Psi&=z\psi(t,z,x)/L~,\\
		A_\mu dx^\mu&=A_t(t,z,x)dt+A_x(t,z,x)dx~.
	\end{align}
	%对于动力学的过程，我们更方便的做法是采用in-goingEddington度规，
	For dynamical processes, it is more convenient to adopt the in-going Eddington metric
	\begin{align}
		ds^{2}=\frac{L^2}{z^2}(-f(z)dt^{2}-2dtdz+dx^{2}+dy^{2})~.
	\end{align}
	The complete set of dynamical evolution equations in one spatial dimension is as follows
	\begin{align}
		2 \lambda  \psi ^3-\alpha  \left(\partial _tA_x\right) \left(\partial _zA_x\right) \psi  z^2 e^{\alpha  \psi ^2 z^2}+2 \partial _t\partial _z\psi +\alpha  \left(\partial _xA_t\right) \left(\partial _zA_x\right) \psi  z^2 e^{\alpha  \psi ^2
			z^2}-\partial _x\partial _x\psi \nonumber\\
		-\frac{1}{2} \alpha  \left(\partial _zA_t)^2\right. \psi  z^2 e^{\alpha  \psi ^2 z^2}+\frac{1}{2} \alpha  \left(\partial _zA_x)^2\right. \psi  z^2 e^{\alpha  \psi ^2 z^2}-\frac{1}{2} \alpha  \left(\partial
		_zA_x)^2\right. \psi  z^5 e^{\alpha  \psi ^2 z^2}\nonumber\\
		+\partial _z\partial _z\psi  z^3-\partial _z\partial _z\psi +3 \left(\partial _z\text{$\psi $)}\right. z^2+3 \tau  \psi ^5 z^2+\psi  z=0&,\label{A1}\\
		%~~~~~~~~~~~~~~~~~~~~~~~~~~~~~~~~~~~~~~~
		-2 \alpha  \left(\partial _x\text{$\psi $)}\right. \left(\partial _zA_x\right) \psi  z^2+2 \alpha  \left(\partial
		_zA_t\right) \psi  z (\left(\partial _z\text{$\psi $)}\right. z+\psi )-\partial _z\partial _xA_x+\partial
		_z\partial _zA_t=0&,\label{A2}\\
		%~~~~~~~~~~~~~~~~~~~~~~~~~~~~~~~~~~~~~~~
		2 \alpha  \left(\partial _tA_x\right) \left(\partial _x\text{$\psi $)}\right. \psi  z^2+\partial _t\partial
		_xA_x+\partial _t\partial _zA_t+2 \alpha  \left(\partial _t\text{$\psi $)}\right. \left(\partial _zA_t\right) \psi 
		z^2-2 \alpha  \left(\partial _xA_t\right) \left(\partial _x\text{$\psi $)}\right. \psi  z^2\nonumber\\
		-\partial _x\partial
		_xA_t+2 \alpha  \left(\partial _x\text{$\psi $)}\right. \left(\partial _zA_x\right) \psi  z^5-2 \alpha 
		\left(\partial _x\text{$\psi $)}\right. \left(\partial _zA_x\right) \psi  z^2+\partial _z\partial _xA_x
		z^3-\partial _z\partial _xA_x=0&,\label{A3}\\
		%~~~~~~~~~~~~~~~~~~~~~~~~~~~~~~~~~~~~~~~
		-2 \alpha  \left(\partial _tA_x\right) \left(\partial _z\text{$\psi $)}\right. \psi  z^2-2 \alpha  \left(\partial
		_tA_x\right) \psi ^2 z-2 \partial _t\partial _zA_x\nonumber\\
		-\left(\partial _zA_x\right) z \left(-2 \alpha  \psi ^2+2 \alpha 
		\left(\partial _t\text{$\psi $)}\right. \psi  z+2 \alpha  \left(\partial _z\text{$\psi $)}\right. \psi 
		\left(z^3-1\right) z+2 \alpha  \psi ^2 z^3+3 z\right)\nonumber\\
		+2 \alpha  \left(\partial _xA_t\right) \psi  z (\left(\partial
		_z\text{$\psi $)}\right. z+\psi )+\partial _z\partial _xA_t-\partial _z\partial _zA_x z^3+\partial _z\partial _zA_x=0&.
		\label{A4}
	\end{align}
	Eq.~(\ref{A3}) is a constraint equation, which in the conformal boundary is
	\begin{align}
		\partial _t\rho=-\partial _x\partial _xA_t-\partial _z\partial _xA_x~.
	\end{align}
	Numerically, we employ the Chebyshev spectral method in the holographic direction with $n_z = 21$ grid points and the Fourier spectral method in the spatial direction. Combined with the Newton iteration, we can solve the equations in the $(z, x)$ plane. For the time evolution, we use the fourth-order Runge–Kutta method with a time step $\delta t = 0.05$.
	
	\subsection{Spontaneous symmetry breaking of $\mathbb{Z}_2$ symmetry}
\begin{figure}
		\centering
		\includegraphics[width=0.32\columnwidth]{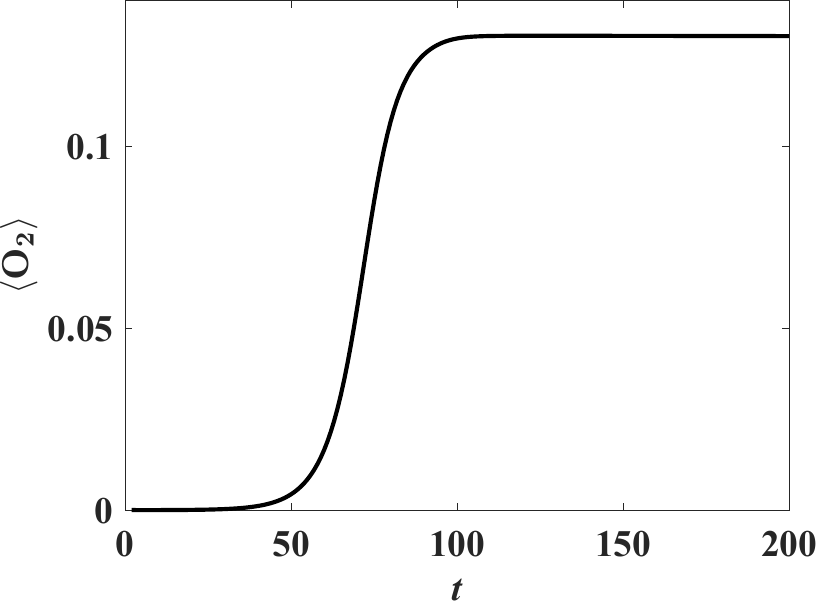}
		\includegraphics[width=0.335\columnwidth]{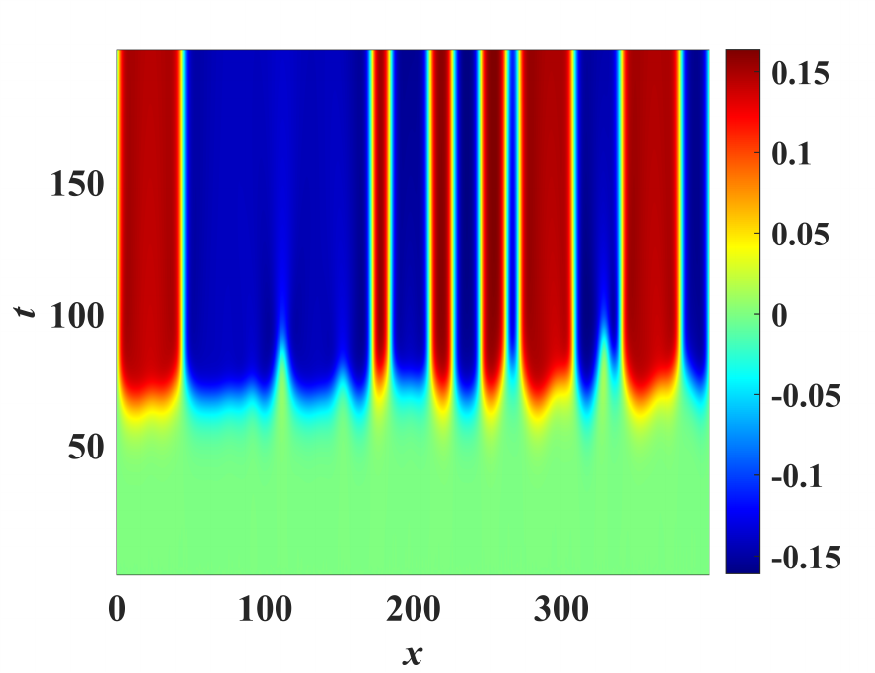}
		\includegraphics[width=0.325\columnwidth]{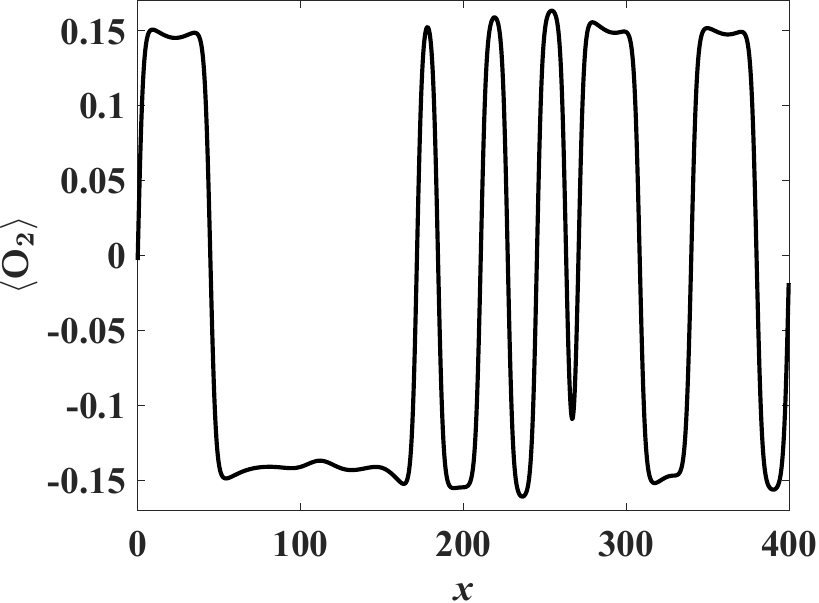}
		\caption{The process of topological defect formation during a quench from $\rho_i = 1.3$ to $\rho_f = 1.52$ in a second-order phase transition (Ref.~\ref{co2nd}) with $L_x=400$, $n_x=1200$, and $\tau_Q=0.1$. The left panel shows the time evolution of the condensate. The middle panel presents a density plot of the condensate as a function of both time and space, where the color bar indicates the magnitude of the condensate. The right panel displays the spatial distribution of the condensate at $t = 200$.}\label{symmetryB}
	\end{figure}

	Typically, if we consider a process of spontaneous symmetry breaking, the initial condition consists of a set of random values. The KZ mechanism \cite{Zeng:2022hut} describes the relationship between the number of topological defects generated during such symmetry breaking and the quench rate. For a system with $\mathbb{Z}_2$ symmetry, it will break to $\psi_+$ or to $\psi_-$. After the symmetry breaking process is complete, the regions connecting $\psi_+$ and $\psi_-$ are kinks. In Fig.~\ref{symmetryB}, we illustrate the process of spontaneous symmetry breaking followed by the formation of a topological defect during a quench across the critical point in a second-order phase transition.

	After the quench ends, the system evolves rapidly. Topological defects appear at approximately the same time, or at very short time intervals. Once topological defects emerge, those that are very close to each other will mutually annihilate. It is worth noting that if the dynamical process here involves only symmetry breaking, then the point with the maximum condensate value in the final spatial structure will be very close to the condensate value of the static solution at the final quench point $\rho_f$.

	\subsection{Phase separation in quench across the critical point}
\begin{figure}
		\centering
		\includegraphics[width=0.31\columnwidth]{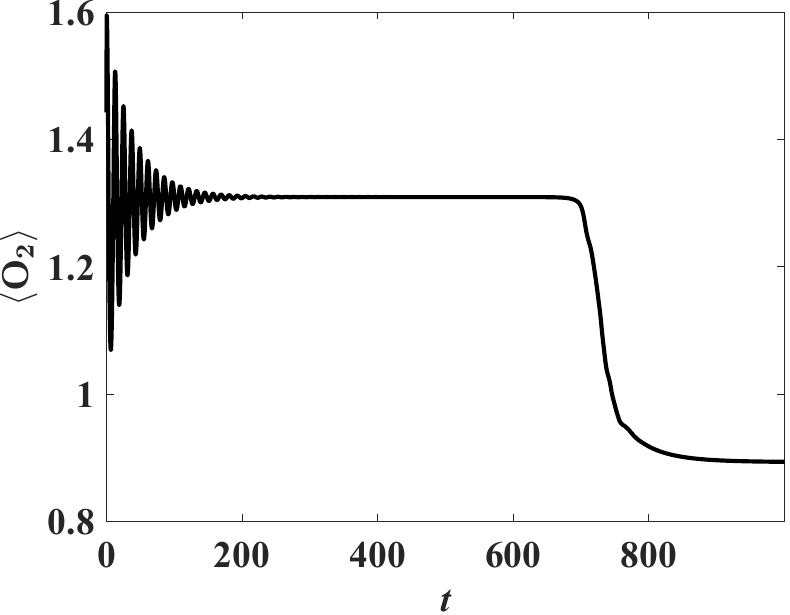}
		\includegraphics[width=0.345\columnwidth]{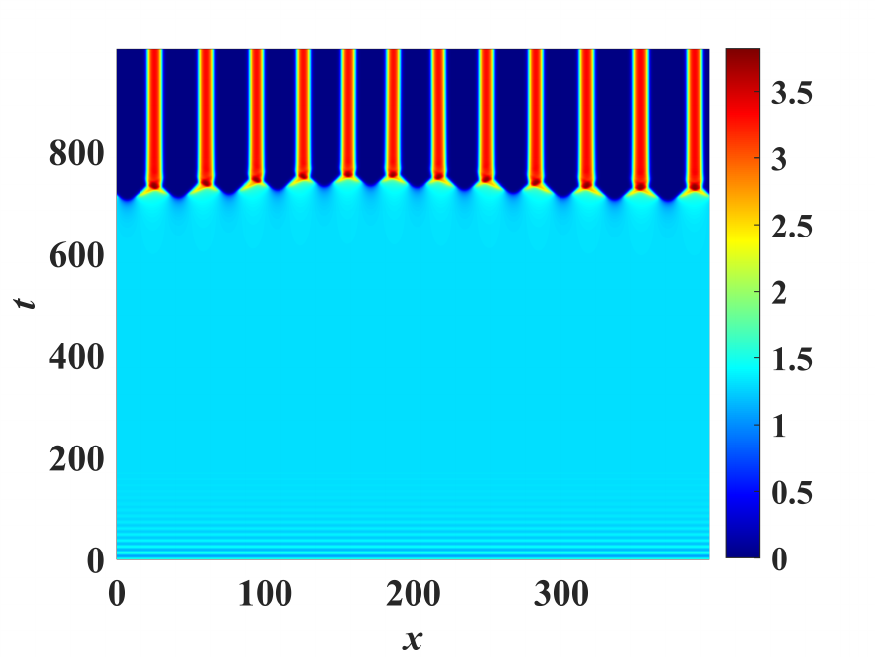}
		\includegraphics[width=0.312\columnwidth]{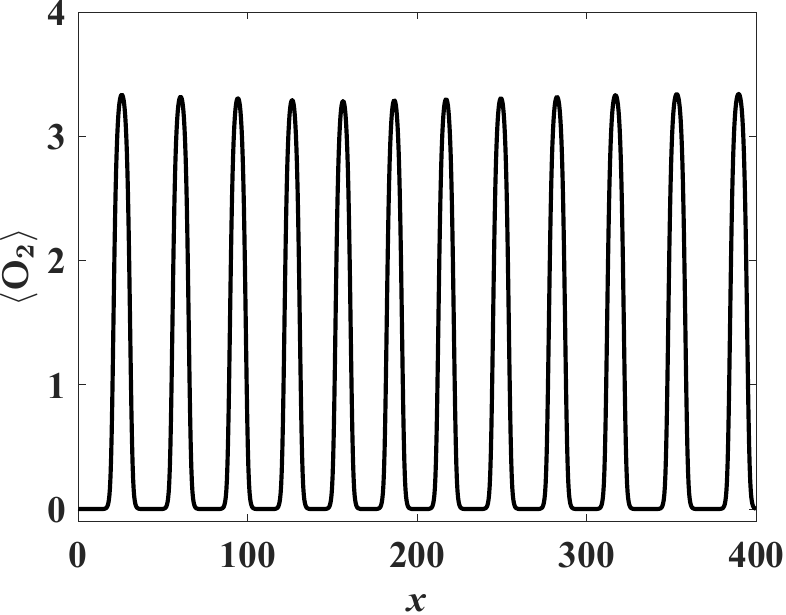}
		\caption{The process of phase separation during a quench from $\rho_i = 1.68$ to $\rho_f = 1.52$ in a first-order phase transition (Ref.~\ref{co1st}) with $L_x=400$, $n_x=1200$, and $\tau_Q=0.1$. The left panel shows the time evolution of the condensate. The middle panel presents a density plot of the condensate as a function of both time and space, where the color bar indicates the magnitude of the condensate. The right panel displays the spatial distribution of the condensate at $t = 1000$.}\label{phaseSp}
	\end{figure}

	We first introduce the process involving only phase separation. When the system parameters lie in the unstable region of a first-order phase transition, quenching the system can induce spatial inhomogeneities even without crossing the critical point. This phenomenon is known as phase separation. In Fig.~\ref{phaseSp}, we show the process of quenching from $\rho_i = 1.68$ to $\rho_f = 1.52$ in a first-order phase transition (Fig.~\ref{co1st}) with $\tau_Q=0.1$. Since the final state lies in the metastable or spinodal decomposition region, the homogeneous initial state becomes unstable, and the system lowers its free energy by spontaneously separating into high-density and low-density regions with different condensate values, eventually forming spatially coexisting bubble structures. This phase separation process is a typical nonequilibrium dynamical phenomenon characterized by the formation and growth of spatial structures. In this quench process involving only phase separation, the final state of the system is macroscopic phase separation driven by thermodynamic instability, rather than topological defects generated by symmetry breaking.
	
\begin{figure}
		\centering
		\includegraphics[width=0.312\columnwidth]{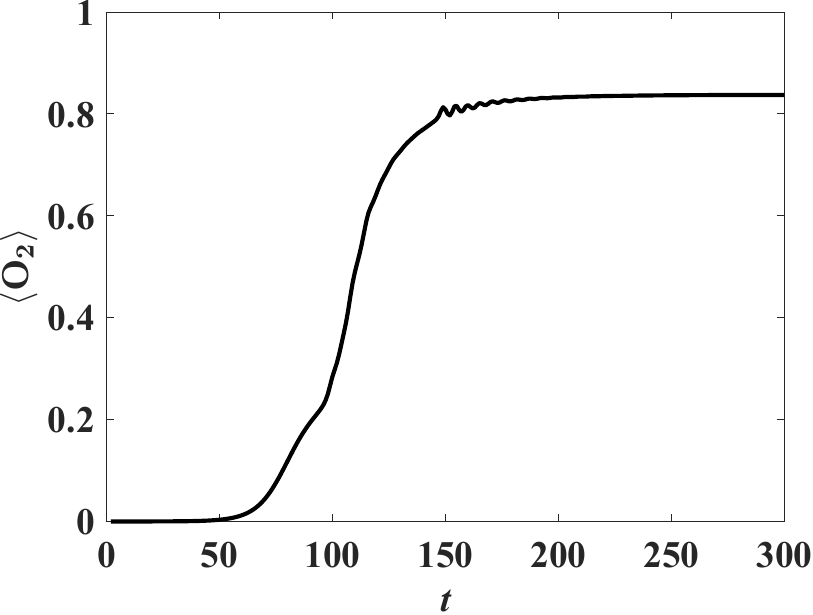}
		\includegraphics[width=0.34\columnwidth]{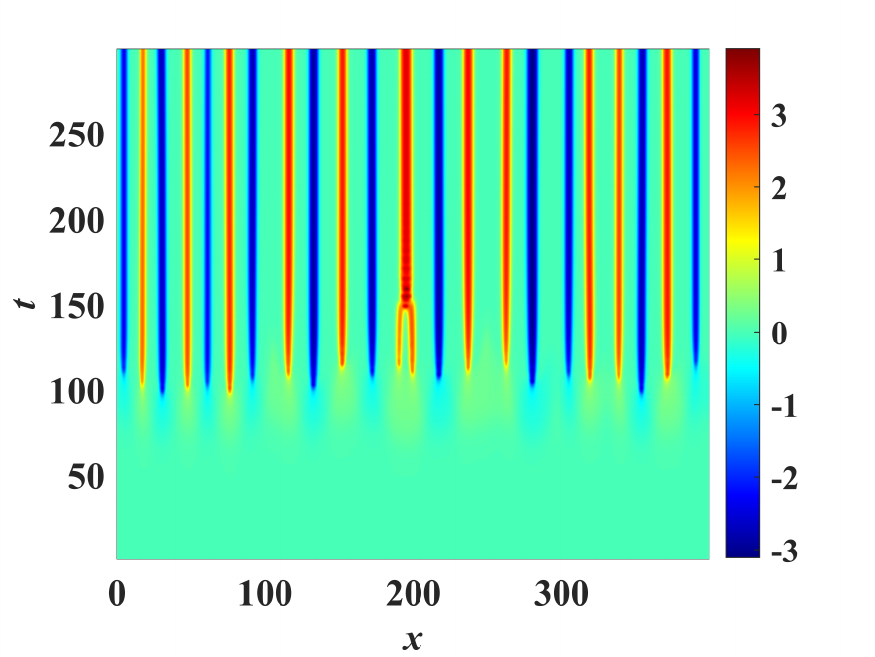}
		\includegraphics[width=0.315\columnwidth]{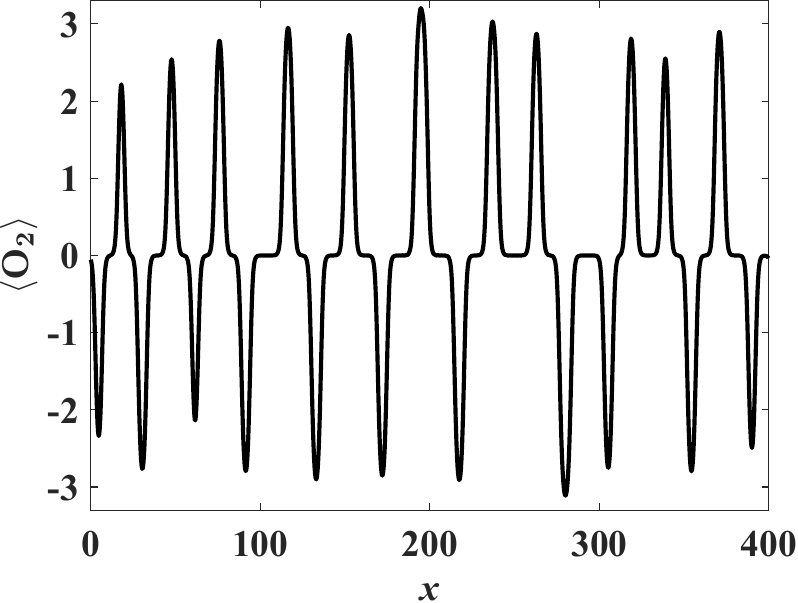}
		\caption{The process of topological defect formation during a quench from $\rho_i = 1.3$ to $\rho_f = 1.52$ in a first-order phase transition (Ref.~\ref{co1st}) with $L_x=400$, $n_x=1200$, and $\tau_Q=0.1$. The left panel shows the time evolution of the condensate. The middle panel presents a density plot of the condensate as a function of both time and space, where the color bar indicates the magnitude of the condensate. The right panel displays the spatial distribution of the condensate at $t = 300$.}\label{symmetryBPS}
	\end{figure}

	If the quench path crosses the critical point and enters the unstable region of a first-order phase transition (e.g., quenching from $\rho_i = 1.3$ to $\rho_f = 1.52$), the system simultaneously undergoes $\mathbb{Z}_2$ symmetry breaking and phase separation, as shown in Fig.~\ref{symmetryBPS}. Compared to the pure phase separation case in Fig.~\ref{phaseSp}, the evolution process and final state in Fig.~\ref{symmetryBPS} exhibit significantly different features. First, in the early stage of evolution, the system rapidly undergoes symmetry breaking, forming multiple kink structures connecting regions of positive and negative condensate values. The large space that would otherwise allow the system to generate multiple bubbles is divided by kinks into many discrete small-scale regions, and the phase separation process can only proceed within these confined small-scale domains. Consequently, in order to maintain the lowest free energy, the original condensate values spontaneously separate into regions with relatively larger and smaller condensate values according to the phase separation mechanism. Therefore, the maximum condensate value in the final stage of evolution is significantly larger than the condensate value of the static solution at the same final quench point, which is the most notable difference between the right panel of Fig.~\ref{symmetryBPS} and that of Fig.~\ref{symmetryB}.
	
	%拓扑缺陷诱导的相分离：扩散现象
	\subsection{Topological defect induced phase separation: invasion phenomenon}
	%在详细介绍了对称性破缺和相分离的机制以及他们的混合效应之后，我们下面将介绍一种特有的现象，叫做入侵现象。
	After introducing in detail the mechanisms of symmetry breaking and phase separation, as well as their combined effects, we will now discuss a unique phenomenon known as the invasion phenomenon \cite{PhysRevE.48.2861,Scheel2017}.

\begin{figure}
		\centering
		\includegraphics[width=0.315\columnwidth]{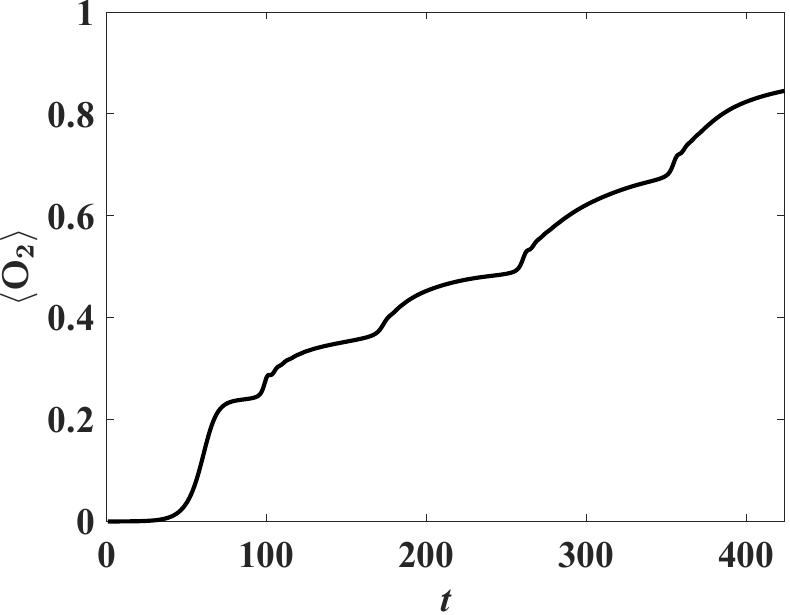}
		\includegraphics[width=0.35\columnwidth]{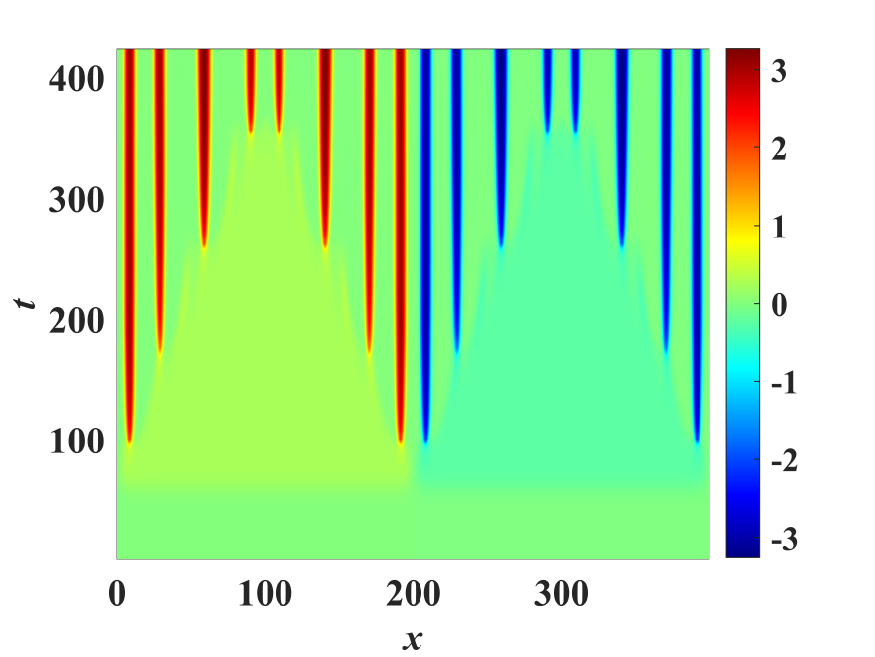}
		\includegraphics[width=0.32\columnwidth]{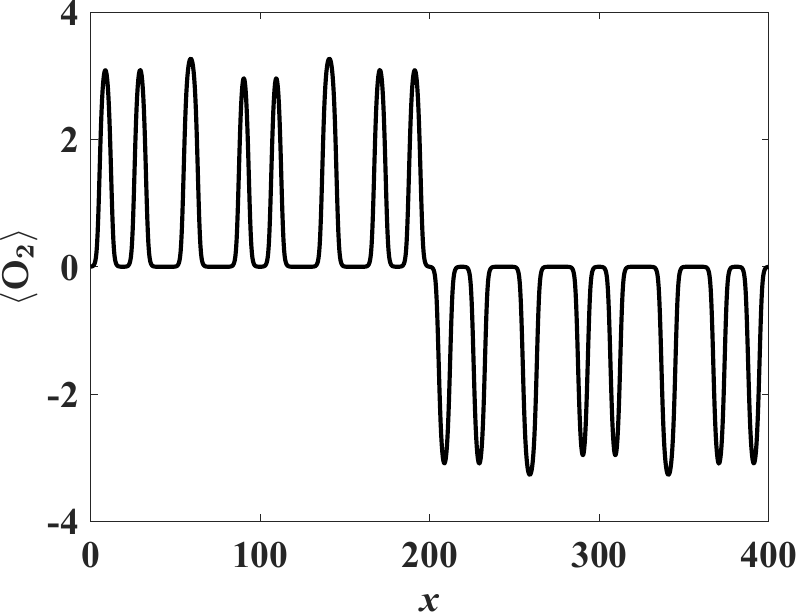}
		\caption{The process of topological defect formation during a quench from $\rho_i = 1.3$ to $\rho_f = 1.52$ in a first-order phase transition (Ref.~\ref{co1st}) with $L_x=400$, $n_x=1200$, and $\tau_Q=0.1$. During this process, the initial condition for $\psi$ is set as $\psi_i(x) = \{10^{-5},0\leq x<L_x/2;-10^{-5},L_x/2<x\leq L_x\}$. The left panel shows the time evolution of the condensate. The middle panel presents a density plot of the condensate as a function of both time and space, where the color bar indicates the magnitude of the condensate. The right panel displays the spatial distribution of the condensate at $t = 420$.}\label{phaseSpinvasion}
	\end{figure}

	In the quench process across the critical point shown in Fig.~\ref{phaseSpinvasion}, we adopt a special initial condition: the entire space is pre-divided into two regions — the left side with a negative condensate value and the right side with a positive condensate value. This initial setup causes the system to form two kink structures at the boundaries between the positive and negative regions in the early stage of evolution. As the quench proceeds, phase separation does not occur simultaneously throughout the entire space but is preferentially triggered at these two kink locations — because the kinks possess the greatest spatial inhomogeneity and dynamically induce inhomogeneous structures first. Subsequently, the phase separation process gradually spreads from these two kinks toward the middle, exhibiting a directional expansion dynamics. This process is called the invasion phenomenon.

	This invasion phenomenon is fundamentally different from bubble nucleation and growth in pure phase separation (Fig.~\ref{phaseSp}). In pure phase separation process, the two phases form interconnected structures through random nucleation without a well-defined starting location. In the invasion process, however, pre-existing kink structures serve as preferential triggering sites for phase separation due to their strong spatial inhomogeneity, endowing the evolution with clear directionality and spatial order. As time progresses, the phase separation fronts emanating from the left and right kinks propagate toward each other and eventually meet in the middle of the two kinks, completing the entire invasion process.
	
	\begin{figure}
		\centering
		\includegraphics[width=0.38\columnwidth]{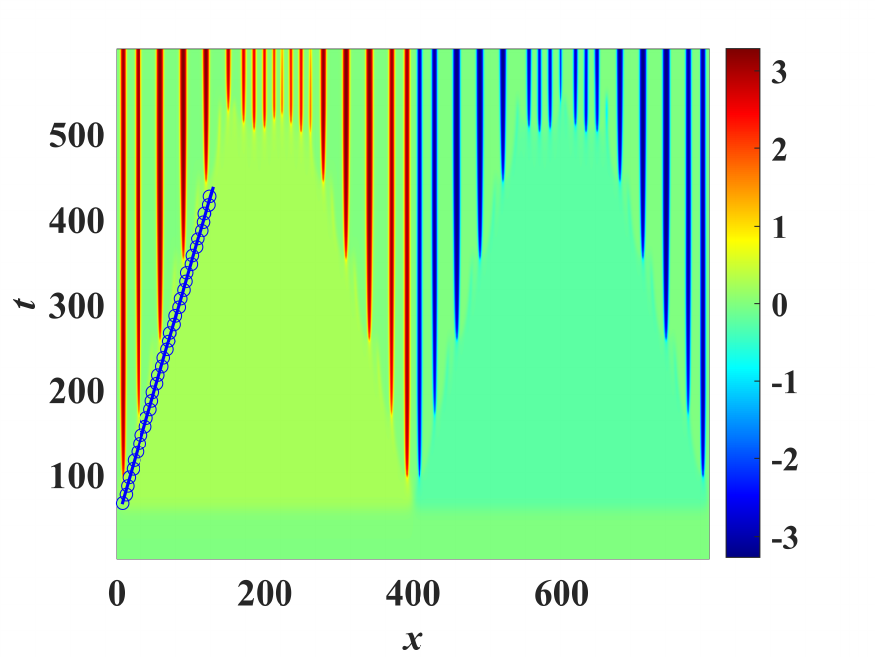}
		\includegraphics[width=0.35\columnwidth]{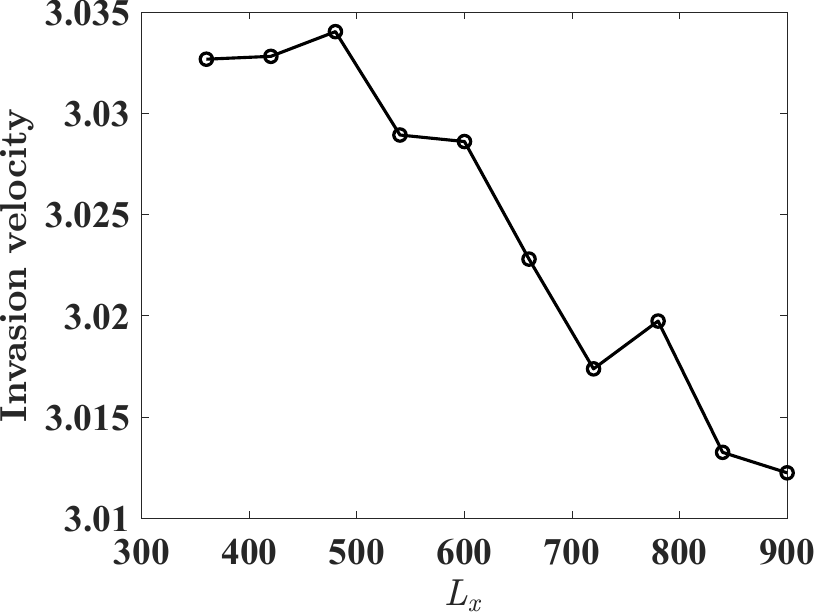}
		\caption{The relationship between the invasion velocity and different spatial scales with a fixed initial configuration. The left panel presents a density plot of the condensate as a function of both time and space for $L_x=800$, where the color bar indicates the magnitude of the condensate. The right panel illustrates the invasion velocity as a function of the spatial scale $L_x$. The blue circles denote the points where the invasion phenomenon occurs at a fixed time. The blue solid line represents the linear fit to these points, with the fitting function given by $f(x) = ax + b$. We define $a$ as the invasion velocity.}\label{LxSpeedand800}
	\end{figure}
	
	%蓝色圆圈是固定时间时发生入侵现象的点，蓝色实线是蓝色圆圈的拟合直线，拟合函数时f（x）=ax+b,我们定义a为入侵速度。
	
	For a fixed final quench point $\rho_f$ and a fixed quench time $\tau_Q$, changing the spatial scale does not alter the invasion velocity. As an example, we show the complete invasion process for $L_x=800$ in the left panel of Fig~\ref{LxSpeedand800}, with the same parameter settings as in Fig.~\ref{phaseSpinvasion}. In Fig.~\ref{LxSpeedand800}, the blue circles indicate the path of the entire invasion process, and the blue solid line represents the linear fit to these points. In this case, we define the slope of this blue solid line as the velocity of the invasion process. In the right panel of Fig.~\ref{LxSpeedand800}, using the same parameter settings but varying the spatial scale, we compute the invasion velocity. The results show that the invasion velocity exhibits negligible variation with changes in spatial scale. In addition, there exists a specific freezing time $t_c$ in the invasion process. When the evolution time exceeds this freezing time $t_c$, the invasion phenomenon disappears and the pure phase separation mechanism governs the evolution of the remaining space. When the total system size is sufficiently small, this freezing phenomenon does not manifest. However, when the system size is large enough, this freezing phenomenon emerges, as shown in the left panel of Fig.~\ref{LxSpeedand800}.
	
	\begin{figure}
		\centering
		\includegraphics[width=0.315\columnwidth]{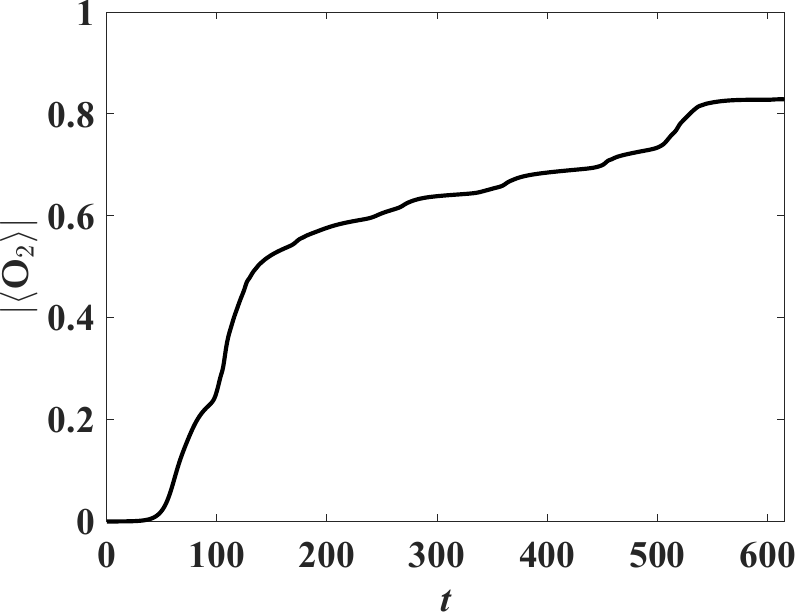}
		\includegraphics[width=0.35\columnwidth]{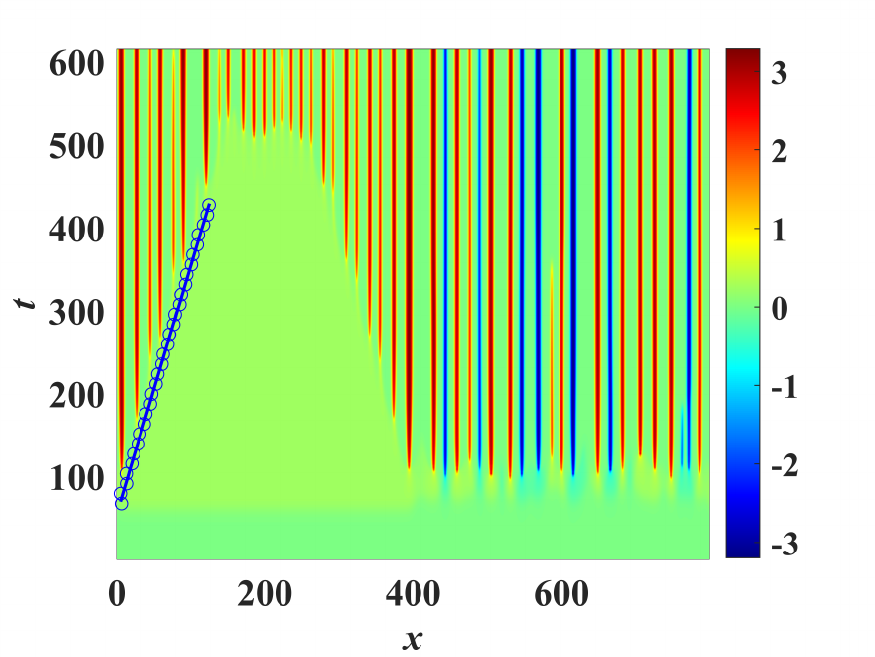}
		\includegraphics[width=0.32\columnwidth]{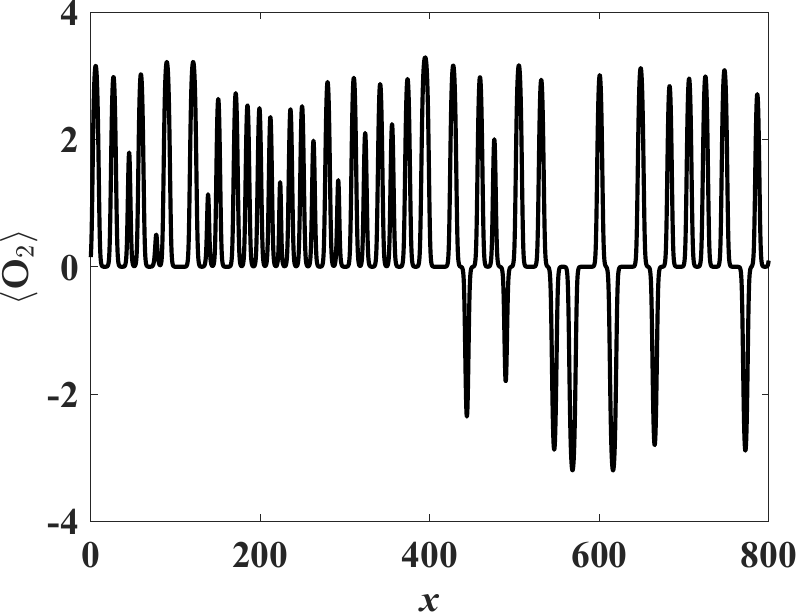}
		\caption{The process of topological defect formation during a quench from $\rho_i = 1.3$ to $\rho_f = 1.52$ in a first-order phase transition (Ref.~\ref{co1st}) with $L_x=800$, $n_x=2400$, and $\tau_Q=0.1$. During this process, the initial condition for $\psi$ is set as $\psi_i(x) = \{10^{-5},0\leq x<L_x/2;10^{-5}\times(random ~seed),L_x/2<x\leq L_x\}$. The left panel shows the time evolution of the condensate. The middle panel presents a density plot of the condensate as a function of both time and space, where the color bar indicates the magnitude of the condensate. The right panel displays the spatial distribution of the condensate at $t = 600$. The blue circles denote the points where the invasion phenomenon occurs at a fixed time. The blue solid line represents the linear fit to these points}\label{phaseSpinvasionHoHy}
	\end{figure}
	
	%最后，让我们来考虑一种非常特殊的情况。我们仍然把整个空间分割成两份，其中一半给固定构型，另外一半给随机初始扰动。这样做的主要目的是为验证随机扰动会不会干扰入侵现象。我们在图[]中给出了相应的结果。在快速quench结束之后，没有给定初始扰动的区域快速发生对称性破缺以及相分离，形成和图6中右图展示的泡泡结构。与之相反的，在固定初始构型的区域，系统仍然展现出了入侵现象，这一现象和图7以及图8中展示的结果一样。尽管图phaseSpinvasionHoHy的入侵过程中相分离效应产生的非均匀结构似乎存在个数上的差异，但是入侵速度是基本上一致的。我们计算了图phaseSpinvasionHoHy的入侵速度，它是a约等于2.99，非常接近图LxSpeedand800中右图展示的结果。这一结果表明，入侵过程不受初始构型的影响，他是一个系统的内禀属性。
	Finally, let us consider a very special case. We again divide the entire space into two regions: one half is assigned a fixed configuration, while the other half is given random initial perturbations. The main purpose of this setup is to verify whether random perturbations interfere with the invasion phenomenon. The corresponding results are presented in Fig.~\ref{phaseSpinvasionHoHy}. After rapid quench, the region without an initial fixed configuration quickly undergoes symmetry breaking and phase separation, forming a bubble structure similar to that shown in the right panel of Fig.~\ref{symmetryBPS}. In contrast, in the region with the fixed initial configuration, the system still exhibits the invasion phenomenon, a result consistent with those shown in Fig.~\ref{phaseSpinvasion} and Fig.~\ref{LxSpeedand800}. Although there appears to be a difference in the number of inhomogeneous structures arising from phase separation during the invasion process in Fig.~\ref{phaseSpinvasionHoHy}, the invasion velocities are essentially consistent. We calculate the invasion velocity in Fig.~\ref{phaseSpinvasionHoHy} to be approximately $a \approx 2.99$, which is very close to the result shown in the right panel of Fig.~\ref{LxSpeedand800}. This indicates that the invasion process is independent of the initial configuration and is an intrinsic property of the system.

	\section{Conclusions and outlooks}\label{sec5}
	%我们在这一章节做一个总结。在本文中，我们研究了具有 $\mathbb{Z}_2$ 对称性的全息超流体模型中对称性破缺与相分离机制耦合动力学过程。通过在标量场势能中引入高阶非线性项 $\lambda\Psi^4$ 和 $\tau\Psi^6$，我们构建了一个包含二级、一级以及COW相变的相图，为研究非平衡现象提供了一个有用的平台。通过对跨越临界点并进入一级相变不稳定区的淬火过程进行系统的数值模拟，我们获得了以下几个关键发现。
	In this paper, we investigate the coupled dynamics of symmetry breaking and phase separation in a holographic superfluid model with $\mathbb{Z}_2$ symmetry. By introducing higher-order nonlinear terms $\lambda\Psi^4$ and $\tau\Psi^6$ into the scalar field potential, we construct a phase diagram that encompasses second-order, first-order, and COW phase transitions, providing a useful platform for studying nonequilibrium phenomena. Through systematic numerical simulations of quench processes that cross the critical point and enter the unstable region of a first-order phase transition, we obtain the following key findings.
	
	%首先，我们证明了当淬火过程同时触发 $\mathbb{Z}_2$ 对称性破缺和相分离时，这两种机制表现出非平凡的耦合效应。对称性破缺过程中产生的拓扑缺陷（kink）将空间分割成受限的区域，相分离随后在这些受限的小尺度区域内进行。这种耦合导致演化终态的凝聚值显著高于相同淬火终点静态解的凝聚值，这是一个区分此混合机制与纯对称性破缺或纯相分离过程的独特标志。
	First, we demonstrate that when a quench simultaneously triggers $\mathbb{Z}_2$ symmetry breaking and phase separation, the two mechanisms exhibit nontrivial coupling. The topological defects (kinks) generated during symmetry breaking divide the spatial domain into confined regions, within which phase separation subsequently proceeds. This coupling leads to final condensate values significantly exceeding those of the static solution at the same quench point, a distinctive signature that distinguishes this mixed regime from pure symmetry breaking or pure phase separation. Second, and most importantly, by preparing initial conditions with well-defined spatial partitions (half positive and half negative condensate values), we realized a special dynamical phenomenon—the invasion process. In this process, pre-existing kink structures serve as preferential triggering sites for phase separation, inducing a directional expansion dynamics that propagates outward from the defects. Under ultrafast quenches, the invasion velocity exhibits spatial scale independence, remaining unchanged as the system size varies. This scale-invariant property suggests that the invasion velocity is an intrinsic characteristic of the coupled dynamics.
	%Finally, we employ the invasion velocity as a dynamical probe to investigate the supercritical region. By systematically varying the final charge density $\rho_f$ in the supercritical region, we find that the invasion velocity exhibits a clear turning point. This turning point defines a new dynamical crossover line in the phase diagram, characterized by nonequilibrium evolution processes, providing a novel criterion for distinguishing different subphases within the supercritical region.

	%未来有几个研究方向值得探索。首先，研究类似的入侵现象是否存在于更高空间维度将是十分有趣的，在那里拓扑缺陷表现为弦或畴壁而非一维的扭结。这些高维缺陷与相分离之间的相互作用可能会产生更丰富的动力学行为。其次，入侵速度与微观参数（如非线性耦合常数 $\lambda$ 和 $\tau$）之间的关系值得系统地探索。理解这种关系可以为缺陷诱导动力学的普适性方面提供更深入的见解。
	%总之，我们的工作研究了对称性破缺和相分离再quench过程中的耦合动力学过程，并且揭示了全息超流体中拓扑缺陷与相分离之间的一种新颖的耦合机制，即入侵现象。这些发现丰富了我们对强耦合系统中非平衡结构形成的理解。
	Several directions for future research are worth exploring. First, it would be interesting to investigate whether similar invasion phenomena exist in higher spatial dimensions, where topological defects take the form of strings or domain walls rather than one-dimensional kinks. The interplay between these higher-dimensional defects and phase separation may yield richer dynamical behaviors. Second, the relationship between the invasion velocity and microscopic parameters, such as the nonlinear coupling constants $\lambda$ and $\tau$, deserves systematic exploration.
	
	In summary, this work investigates the coupled dynamics of symmetry breaking and phase separation during a quench process, and reveals a novel coupling mechanism between topological defects and phase separation in holographic superfluids—namely, the invasion phenomenon. These findings enrich our understanding of nonequilibrium structure formation in strongly coupled systems.
	% In summary, our work reveals a novel coupling mechanism between topological defects and phase separation in holographic superfluids, namely the invasion phenomenon
	% %, and demonstrates its utility as a new criterion for identifying crossovers within the supercritical region. 
	% These findings enrich our understanding of nonequilibrium structure formation in strongly coupled systems.

	\section*{Acknowledgements}
	This work is partially supported by NSFC with Grant Nos. 12575054, 12533001, 12575049, 12473001, 12205039, 12305058 and 11965013. ZYN is partially supported by Yunnan High-level Talent Training Support Plan Young $\&$ Elite Talents Project (Grant No. YNWR-QNBJ-2018-181). This work is also supported by the National SKA Program of China (Grant Nos. 2022SKA0110200 and 2022SKA0110203) and the National 111 Project (Grant No. B16009).

	\bibliographystyle{JHEP}
	\bibliography{reference}
\end{document}